# Stochastic Database Cracking: Towards Robust Adaptive Indexing in Main-Memory Column-Stores[*]

Felix Halim⋆    Stratos Idreos†    Panagiotis Karras⋄    Roland H. C. Yap⋆

⋆National University of Singapore    †CWI, Amsterdam    ⋄Rutgers University
{halim, ryap}@comp.nus.edu.sg    idreos@cwi.nl    karras@business.rutgers.edu

## ABSTRACT

Modern business applications and scientific databases call for inherently dynamic data storage environments. Such environments are characterized by two challenging features: (a) they have little idle system time to devote on physical design; and (b) there is little, if any, a priori workload knowledge, while the query and data workload keeps changing dynamically. In such environments, traditional approaches to index building and maintenance cannot apply. *Database cracking* has been proposed as a solution that allows on-the-fly physical data reorganization, as a collateral effect of query processing. Cracking aims to continuously and automatically adapt indexes to the workload at hand, without human intervention. Indexes are built incrementally, adaptively, and on demand. Nevertheless, as we show, existing adaptive indexing methods fail to deliver *workload-robustness*; they perform much better with random workloads than with others. This frailty derives from the inelasticity with which these approaches interpret each query as a hint on how data should be stored. Current cracking schemes *blindly* reorganize the data within each query's range, even if that results into successive expensive operations with minimal indexing benefit.

In this paper, we introduce *stochastic cracking*, a significantly more resilient approach to adaptive indexing. Stochastic cracking also uses each query as a hint on how to reorganize data, but not blindly so; it gains resilience and avoids performance bottlenecks by deliberately applying certain arbitrary choices in its decision-making. Thereby, we bring adaptive indexing forward to a mature formulation that confers the workload-robustness previous approaches lacked. Our extensive experimental study verifies that stochastic cracking maintains the desired properties of original database cracking while at the same time it performs well with diverse realistic workloads.

## 1. INTRODUCTION

Database research has set out to reexamine established assumptions in order to meet the new challenges posed by big data, scientific databases, highly dynamic, distributed, and multi-core CPU environments. One of the major challenges is to create simple-to-use and flexible database systems that have the ability self-organize according to the environment [7].

**Physical Design.** Good performance in database systems largely relies on proper *tuning* and *physical design*. Typically, all tuning choices happen up front, assuming sufficient workload knowledge and idle time. Workload knowledge is necessary in order to determine the appropriate tuning actions, while idle time is required in order to perform those actions. Modern database systems rely on auto-tuning tools to carry out these steps, e.g., [6, 8, 13, 1, 28].

**Dynamic Environments.** However, in dynamic environments, workload knowledge and idle time are scarce resources. For example, in scientific databases new data arrives on a daily or even hourly basis, while query patterns follow an exploratory path as the scientists try to interpret the data and understand the patterns observed; there is no time and knowledge to analyze and prepare a different physical design every hour or even every day.

Traditional indexing presents three fundamental weaknesses in such cases: (a) the workload may have changed by the time we finish tuning; (b) there may be no time to finish tuning properly; and (c) there is no indexing support during tuning.

**Database Cracking.** Recently, a new approach to the physical design problem was proposed, namely *database cracking* [14]. Cracking introduces the notion of continuous, incremental, partial and on demand adaptive indexing. Thereby, indexes are incrementally built and refined during query processing. Cracking was proposed in the context of modern column-stores and has been hitherto applied for boosting the performance of the select operator [16], maintenance under updates [17], and arbitrary multi-attribute queries [18]. In addition, more recently these ideas have been extended to exploit a partition/merge -like logic [19, 11, 12].

**Workload Robustness.** Nevertheless, existing cracking schemes have not deeply questioned the particular *way* in which they interpret queries as a hint on how to organize the data store. They have adopted a simple interpretation, in which a select operator is taken to describe a range of the data that a *discriminative* cracker index should provide easy access to for future queries; the remainder of the data remains non-indexed until a query expresses interest therein. This simplicity confers advantages such as *instant and lightweight adaptation*; still, as we show, it also creates a problem.

Existing cracking schemes faithfully and obediently follow the hints provided by the queries in a workload, without examining whether these hints make good sense from a broader view. This approach fares quite well with random workloads, or workloads that expose consistent interest in certain regions of the data. However, in other realistic workloads, this approach can falter. For example, consider a workload where successive queries ask for consecutive items, as if they sequentially scan the value domain; we call this workload pattern *sequential*. Applying existing cracking methods on this workload would result into repeatedly reorganizing large chunks of data with every query; yet this expensive operation confers only a minor benefit to subsequent queries. Thus, existing cracking schemes fail in terms of *workload robustness*.

---

[*]Work supported by Singapore's MOE AcRF grant T1 251RES0807.

Such a workload robustness problem emerges with any workload that focuses in a specific area of the value domain at a time, leaving (large) unindexed data pieces that can cause performance degradation if queries touch this area later on. Such workloads occur in exploratory settings; for example, in scientific data analysis in the astronomy domain, scientists typically "scan" one part of the sky at a time through the images downloaded from telescopes.

A natural question regarding such workloads is whether we can anticipate such access patterns in advance; if that were the case, we would know what kind of indexes we need, and adaptive indexing techniques would not be required. However, this may not always be the case; in exploratory scenarios, the next query or the next batch of queries typically depends on the kind of answers the user got for the previous queries. Even in cases where a pattern can be anticipated, the benefits of adaptive indexing still apply, as it allows for straightforward access to the data without the overhead of a priori indexing. As we will see in experiments with the data and queries from the Sloan Digital Sky Survey/SkyServer, by the time full indexing is still partway towards preparing a traditional full index, an adaptive indexing technique will have already answered $1.6 * 10^5$ queries. Thus, in exploratory scenarios such as scientific databases [15, 20], it is critical to assure such a *quick* gateway to the data in a robust way that works with *any* kind of workload.

Overall, the workload robustness requirement is a major challenge for future database systems [9]. While we know how to build well-performing specialized systems, designing systems that perform well over a broad range of scenarios and environments is significantly harder. We emphasize that this workload robustness imperative does not imply that a system should perform all conceivable tasks efficiently; it is accepted nowadays that "one size does not fit all" [26]. However, it does imply that a system's performance should not deteriorate after changing a minor detail in its input or environment specifications. The system should maintain its performance and properties when faced with such changes. The whole spectrum of database design and architecture should be re-investigated with workload robustness in mind [9], including, e.g., optimizer policies and low-level operator design.

**Contributions.** In this paper, we design cracking schemes that satisfy the workload-robustness imperative. To do so, we reexamine the underlying assumptions of existing schemes and propose a significantly more resilient alternative. We show that original cracking relies on the *randomness* of the workloads to converge well; we argue that, to succeed with non-random workloads, cracking needs to introduce randomness on its own. Our proposal introduces arbitrary and random, or *stochastic*, elements in the cracking process; each query is still taken as a hint on how to reorganize the data, albeit in a *lax* manner that allows for reorganization steps not explicitly dictated by the query itself. While we introduce such auxiliary actions, we also need to maintain the lightweight character of existing cracking schemes. To contain the overhead brought about by stochastic operations, we introduce *progressive* cracking, in which a single cracking action is completed *collaboratively* by multiple queries instead of a single one. Our experimental study shows that stochastic cracking preserves the benefits of original cracking schemes, while also expanding these benefits to a large variety of realistic workloads on which original cracking fails.

**Organization.** Section 2 provides an overview of related work and database cracking. Then, Section 3 motivates the problem through a detailed evaluation of original cracking, exposing its weaknesses under certain workloads. Section 4 introduces stochastic cracking, while Section 5 presents a thorough experimental analysis. Sections 6 and 7 discuss future work and conclude the paper.

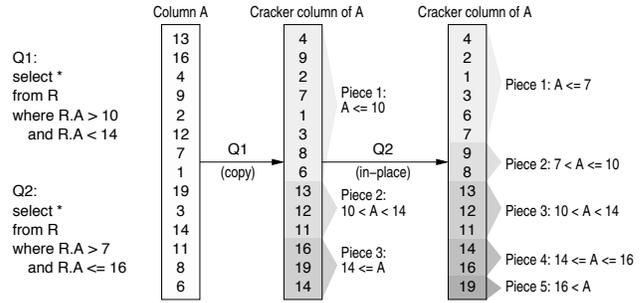

**Figure 1: Cracking a column.**

## 2. RELATED WORK

Here, we briefly recap three approaches to indexing and tuning: offline analysis, online analysis, and the novel cracking approach.

**Offline Analysis.** Offline analysis or *auto-tuning* tools exist in every major database product. They rely on the what-if analysis paradigm and close interaction with the system's query optimizer [6, 8, 13, 1, 28]. Such approaches are non-adaptive: they render index tuning distinct from query processing operations. They first monitor a running workload and then decide what indexes to create or drop based on the observed patterns. Once a decision is made, it affects all key ranges in an index, while index tuning and creation costs impact the database workload as well. Unfortunately, one may not have sufficient workload knowledge and/or idle time to invest in offline analysis in the first place. Furthermore, with dynamic workloads, any offline decision may soon become invalid.

**Online Analysis.** Online analysis aims to tackle the problem posed by such dynamic workloads. A number of recent efforts attempt to provide viable online indexing solutions [5, 24, 4, 21]. Their main common idea is to apply the basic concepts of offline analysis online: the system monitors its workload and performance while processing queries, probes the need for different indexes and, once certain thresholds are passed, triggers the creation of such new indexes and possibly drops old ones. However, online analysis may severely overload individual query processing during index creation. Approaches such as soft indexes [21] try to exploit the scan of relevant data (e.g., by a select operator) and send this data to a full-index creation routine at the same time. This way, data to be indexed is read only once. Still, the problem remains that creating *full* indexes significantly penalizes individual queries.

**Database Cracking.** The drawbacks of offline and online analysis motivate *adaptive indexing*, the prime example of which is *database cracking* [14]. Database cracking pioneered the notion of continuously and incrementally building and refining indexes as part of query processing; it enables efficient adaptive indexing, where index creation and optimization occur collaterally to query execution; thus, only those tables, columns, and key ranges that are queried are being optimized. The more often a key range is queried, the more its representation is optimized. Non-queried columns remain non-indexed, and non-queried key ranges are not optimized.

**Selection Cracking.** We now briefly recap *selection cracking* [16]. The main innovation is that the physical data store is continuously changing with each incoming query $q$, using *q as a hint on how data should be stored*. Assume a query requests $A<10$. In response, a cracking DBMS clusters all tuples of $A$ with $A<10$ at the beginning of the respective column $C$, while pushing all tuples with $A \geq 10$ to the end. A subsequent query requesting $A \geq v_1$, where $v_1 \geq 10$, has to search and crack only the last part of $C$ where values $A \geq 10$ reside. Likewise, a query that requests $A < v_2$, where $v_2 \leq 10$, searches and cracks only the first part of $C$. All crack actions happen as part of the query operators, requiring no external administration. Figure 1 shows an example of two queries cracking a column using their selection predicates as the partitioning bounds. Query Q1 cuts the column in three pieces and then Q2 enhances this partitioning more by cutting the first and the last piece even further, i.e., where its low and high bound fall. Each query has collected its qualifying tuples in a contiguous area.

Cracking gradually improves data access, eventually leading to a significant speed-up in query processing [16, 18], even during updates [17]; as it is designed over a column-store it is applied at the attribute level; a query results in reorganizing the referenced column(s), not the complete table; it is propagated across multiple columns on demand, depending on query needs with *partial sideways cracking* [18], whereby pieces of cracker columns are dynamically created and deleted based on storage restrictions.

*Adaptive merging* [11, 12], extends cracking to adopt a partition/merge-like logic with active sorting steps; while original cracking can be seen as an incremental quicksort, adaptive merging can be seen as an incremental external merge sort. More recently, [19] studied the broader space of adaptive indexing; it combines insights from both cracking [16] and adaptive merging [11, 12], to devise adaptive indexing algorithms (from very active to very lazy) that improve over both these predecessors.

The benchmark proposed in [10] discusses the requirements for adaptive indexing; (a) lightweight initialization, i.e., low cost for the first few queries that trigger adaptation; and (b) as fast as possible convergence to the desired performance. Initialization cost is measured against that of a full scan, while desired performance is measured against that of a full index. A good adaptive indexing technique should strike a balance between those two conflicting parameters [10, 19]. We follow these guidelines in this paper as well.

To date, all work on cracking and adaptive indexing has focused on main memory environments; persistent data may be on disk but the working data set for a given query (operator in a column-store) should fit in memory for efficient query processing. In addition, the partition/merge-like logic introduced in [19, 11, 12] can be exploited for external cracking.

The basic underlying physical reorganization routines remain unchanged in all cracking work; therefore, for ease of presentation, we develop our work building on the original cracking example. In Section 5, we show that its effect remains the same in the more recent adaptive indexing variants [19].

**Column-Stores.** Database cracking relies on a number of modern column-store design characteristics. Column-stores store data one column at a time in fixed-width dense arrays [22, 27, 3]. This representation is the same both on disk and in memory and allows for efficient physical reorganization of arrays. Similarly, column-stores rely on bulk and vector-wise processing. Thus, a select operator typically processes a single column in vector format at once, instead of whole tuples one at a time. In effect, cracking performs all physical reorganization actions efficiently in one go over a column. For example, the cracking select operator physically reorganizes the proper pieces of a column to bring all qualifying values in a contiguous area and then returns a view of this area as the result.

## 3. THE PROBLEM

In this section, we analyze the properties of database cracking and its performance features. We demonstrate its adaptation potential but also its workload robustness deficiency.

**Cracking Features.** The main power of original database cracking is its ability to self-organize automatically and at low cost.

The former feature (automatic self-organization) is crucial because with automatic adaptation, no special decision-making is required as to when the system should perform self-organizing actions; the system self-organizes *continuously* by default. Apart from conferring benefits of efficiency and administrator-free convenience in workload analysis, automatic self-organization also brings instant, online adaptation in response to a changing workload, without delays. In effect, there is no performance penalty due to having an unsuitable physical design for a prolonged time period.

The latter feature (low cost) is also a powerful property that sets cracking apart from approaches such as online indexing. This property comes from the ability to provide incremental and partial indexing integrated in an efficient way *inside* the database kernel.

**Cracking Continuous Adaptation.** As we have seen in the example of Figure 1, cracking feeds from the select operator, using the selection predicates to drive the way data is stored. After each query, data is clustered in a way such that the qualifying values for the respective select operator are in a contiguous area. The more the queries processed, the more the knowledge and structure introduced; thus, cracking continuously adapts to the workload.

**Cracking Cost.** Let us now discuss the cost of cracking, i.e., the cost to run the select operator, which includes the cost of identifying what kind of physical reorganizations are needed and performing such reorganizations.

A cracking DBMS maintains indexes showing which piece holds which value range, in a tree structure; original cracking uses AVL-trees [16]. These trees are meant to maintain small depth by restricting the number of entries (or the minimum size of a cracking piece); thus, the cost of reorganizing data becomes the dominant part of the whole cracking cost. We can concretely identify this cost as the amount of data the system has to touch for every query, i.e., the number of tuples cracking has to analyze during a select operator. For example, in Figure 1 $Q1$ needs to analyze all tuples in the column in order to achieve the initial clustering, as there is no prior knowledge about the structure of the data. The second query, $Q2$, can exploit the knowledge gained by $Q1$ and avoid touching part of the data. With $Q1$ having already clustered the data into three pieces, $Q2$ needs to touch only two of those, namely the first and third piece. That is because the second piece created by $Q1$ already qualifies for $Q2$ as well.

Generalizing the above analysis, we infer that, with such range queries (select operators), cracking needs to analyze at most two (*end*) pieces per query, i.e., the ones intersecting with the query's value range boundaries. As more pieces are created by every query that does not find an exact match, pieces become smaller.

**Basic Cracking Performance.** Figure 2(a) shows a performance example where cracking (Crack) is compared against a full indexing approach (Sort), in which we completely sort the column with the first query. The data consists of $10^8$ tuples of unique integers, while the query workload is completely random (the ranges requested have a fixed selectivity of 10 tuples per query but the actual bounds requested are random). This scenario assumes a dynamic environment where there is no workload knowledge or idle time in order to pre-sort the data, i.e., our very motivating example for adaptive indexing. As Figure 2(a) shows, once the data is sorted with the first query, from then on performance is extremely fast as we only need to perform a binary search over the sorted column to satisfy each select operator request. Nevertheless, the problem is that we overload the first query. On the other hand, Crack continuously improves performance without penalizing individual queries. Eventually, its performance reaches the levels of

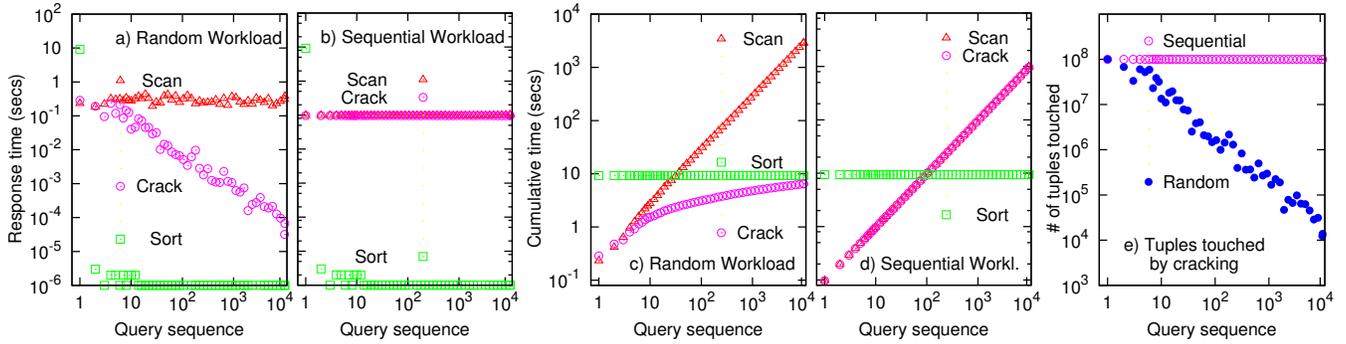

**Figure 2: Basic cracking performance. Per query costs (a,b). Cumulative costs (c,d). Tuples touched (e).**

Sort. We also compare against a plain Scan approach where data is always completely scanned. Naturally, this has a stable behavior; interestingly, Crack does not significantly penalize any query more than the default Scan approach. We emphasize that while Crack and Sort can simply return a view of the (contiguous) qualifying tuples, Scan has to materialize a new array with the result.

**Ideal Cracking Costs.** The ideal performance comes when analyzing fewer tuples. Such a disposition is workload-dependent; it depends not only on the nature of queries posed but also on the *order* in which they are posed. As in the analysis of the quicksort algorithm, Crack achieves the best-case performance (assuming a full column is relevant for the total workload) if each query cracks a piece of the column in exactly two half pieces: the first query splits the column in two equally sized pieces; the second and third query split it in four equal pieces, and so on, resulting in a uniform clustering of the data and gradual improvement of access patterns.

**A Non-ideal Workload.** If we relax the above ideal workload assumption and consider arbitrary query sequences, it is easy to see that, in the general case, the same cumulative pattern holds; the more queries we have seen in the past, the more chances we have to improve performance in the future, as we keep adding knowledge to the crack columns regardless of the exact query pattern. However, the *rate* at which performance improves crucially depends on the query pattern and order. Depending on the query sequence, performance might improve more quickly or slowly in terms of the number of queries needed to achieve a certain performance level.

Let us give a characteristic example through a specific realistic workload. Assume what we call a *sequential* workload, i.e., a workload where every query requests a range which follows the preceding one in a sequence. Say the value domain for a given column $A$ is $[0, 100]$, and the first query requests $A < 1$, the second query requests $A < 2$, the third $A < 3$, and so on. Figure 7 shows an example of such a sequential workload among many others. If we assume that the column has $N$ tuples with unique integers, then the first query will cost $N$ comparisons, the second query will cost $N-1$, the third $N-2$ and so on, causing such a workload to exhibit a very slow adaptation rate. By contrast, in the ideal case where the first query splits the column into two equal parts, the second query already had a reduced cost of $N/2$ comparisons.

Figure 2(b) shows the results with such a workload. As in Figure 2(a), we test Crack against Scan and Sort. The setup is exactly the same as before, i.e., the data in the column, the initial status, and the query selectivity are the same as in the experiment for Figure 2(a); the only difference is that this time queries follow the sequential workload. We observe that Sort and Scan are not affected by the kind of workload tested; their behavior with random and sequential workloads do not deviate significantly from each other. This is not surprising, as the Scan will always scan $N$ tuples no matter the workload, while the full indexing approach will always pay for the complete sort with the first query and then exploit binary search. A slight improvement observed in the Scan performance is due to the short-circuiting in the if statement checking for the requested range. Likewise, there is slight improvement for the Sort strategy after the first query due to caching effects of the binary search in successive short ranges. By contrast, Figure 2(b) clearly shows that Crack fails to deliver the performance improvements seen for the random workload in Figure 2(a). Now its performance does not outperform that of Scan, whereas with the random workload performance improved significantly already after a handful of queries.

To elaborate on this result, Figure 2(e) shows the number of tuples each cracking query needs to touch with these two workloads. With the sequential workload, Crack touches a large number of tuples, which falls only negligibly as new queries arrive, whereas with the random workload the number of touched tuples drops swiftly after only a few queries. With less data to analyze, performance improves rapidly.

Figures 2(c) and (d) present the results of the same two experiments using a different metric, i.e., cumulative response time. Significantly, with the random workload, even after $10^4$ queries, Sort has still not amortized its initialization overhead over Crack. This result shows the principal advantage of database cracking: its lightweight adaptation. However, once we move to the sequential workload, this key benefit is lost; for the first several thousand queries Crack behaves quite similarly to Scan, while Sort amortizes its initialization cost after only 100 queries.

To sum up, while original cracking gives excellent adaptive performance with a random workload, it can at best match the performance of Scan with a pathological, yet realistic, workload.

## 4. STOCHASTIC CRACKING

Having discussed the problem, we now present our proposal in a series of incrementally more sophisticated algorithms that aim to achieve the desired workload robustness while maintaining the adaptability of existing cracking schemes.

**The Source of the Problem.** In Section 3, we have shown that the cost of a query (select operator) with cracking depends on the amount of data that needs to be analyzed for physical reorganization. The sequential workload which we have used as an example to demonstrate the weakness of original cracking, forces cracking to repeatedly analyze large data portions for consecutive queries.

This effect is due to the fact that cracking treats each query as a hint on how to reorganize data in a *blinkered* manner: it takes each query as a literal instruction on what data to index, without looking at the bigger picture. It is thanks to this literalness that cracking can instantly adapt to a random workload; yet, as we have shown, this literal character can also be a liability. With a non-ideal workload, strictly adhering to the queries and reorganizing the array so as to collect the query result, and only that, in a contiguous area, amounts to an inefficient quicksort-like operation; small successive portions of the array are clustered, one after the other, while leaving the rest of the array unaffected. Each new query, having a bound inside the unindexed area of the array, reanalyzes this area all over again.

**The Source of the Solution.** To address this problem, we venture to drop the strict requirement in original cracking that each individual query be literally interpreted as a re-organization suggestion. Instead, we want to force reorganization actions that are not strictly driven by what a query requests, but are still beneficial for the workload at large.

To achieve this outcome, we propose that reorganization actions be *partially* driven by what queries want, and *partially* arbitrary in character. We name the resulting cracking variant *stochastic*, in order to indicate the arbitrary nature of some of its reorganization actions. We emphasize that our new variant should not totally forgo the *query-driven* character of original cracking. An extreme stochastic cracking implementation could adopt a totally arbitrary approach, making random reorganizations along with each query (we discuss such naive cracking variants in Section 5). However, such an approach would discard a feature of cracking that is worth keeping, namely the capacity to adapt to a workload without significant delays. Besides, as we have seen in Figure 2(a), cracking barely imposes any overhead over the default scan approach; while the system adapts, users do not notice significantly slower response times; they just observe faster reaction times later. Our solution should maintain this lightweight property of original cracking too.

Our solution is a sophisticated intermediary between totally query-driven and totally arbitrary reorganization steps performed with each query. It maintains the lightweight and adaptive character of existing cracking, while extending its applicability to practically any workload. In the rest of this section, we present techniques that try to strike a balance between (a) adding auxiliary reorganization steps with each query, and (b) remaining lightweight enough so as not to significantly (if at all) penalize individual queries.

**Stochastic Cracking Algorithms.** All our algorithms are proposed as replacements for the original cracking physical reorganization algorithm [16]. From a high level point of view, nothing changes, i.e., stochastic cracking maintains the design principles for cracking a column-store. As in original cracking [16], in stochastic cracking the select operator physically reorganizes an array that represents a single attribute in a column-store so as to introduce range partitioning information. Meanwhile, a tree structure maintains structural knowledge, i.e., keeps track of which piece of the clustered array contains which value range. As new queries arrive, the select operators therein trigger cracking actions. Each select operator requests for a range of values on a given attribute (array) and the system reacts by physically reorganizing this array, if necessary, and collecting all qualifying tuples in a continuous area. The difference we introduce with stochastic cracking is that, instead of passively relying on the workload to stipulate the *kind* and *timing* of reorganizations taking place, it exercises more control over these decisions.

**Algorithm DDC.** Our first algorithm, the *Data Driven Center* algorithm (DDC), exercises its own decision-making without using random elements; we use it as a baseline for the subsequent development of its genuinely stochastic variants. The motivation for DDC comes from our analysis of the ideal cracking behavior in Section 3; ideally, each reorganization action should split the respective array piece in *half*, in a quicksort-like fashion. DDC *recursively* halves relevant pieces on its way to the requested range, introducing several new pieces with each new query, especially for the first queries that touch a given column. The term "*Center*" in its name denotes that it always tries to cut pieces in half.

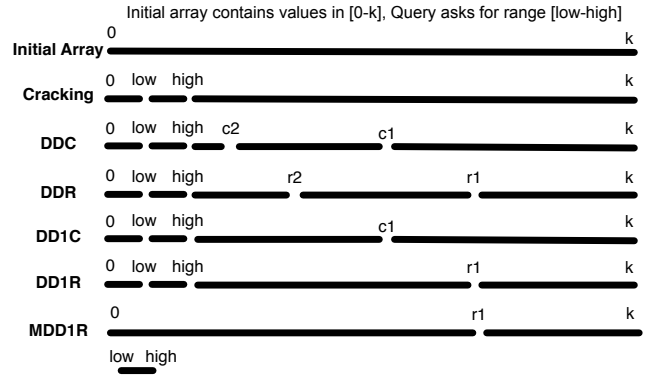

**Figure 3: Cracking algorithms in action.**

The other component in its name, namely "*Data Driven*", contrasts it to the *query-driven* character of default cracking; if a query requests the range $[a, b]$, default cracking reorganizes the array based on $[a, b]$ regardless of the actual data. By contrast, DDC takes the data into account. Regardless of what kind of query arrives, DDC always performs specific data-driven actions, *in addition* to query-driven actions. The query-driven mentality is maintained, as otherwise the algorithm would not provide good adaptation.

Given a query in $[a, b]$, DDC *recursively* halves the array piece where $[a, b]$ falls, until it reaches a point where the size of the resulting piece is sufficiently small. Then, it cracks this piece based on $[a, b]$. As with original cracking, a request for $[a, b]$ in an already cracked column will in general result in two requests/cracks; one for $[a, )$ and one for $(, b]$ (as for Q2 in Fig. 1).

A high-level example for DDC is given in Figure 3. This figure shows the end result of a simplifying example of data reorganization with the various stochastic cracking algorithms that we introduce, as well as with original cracking. An array, initially uncracked, is queried for a value range in [low, high]. The initially uncracked array, as well as the separate pieces created by the various cracking algorithms, are represented by continuous lines. We emphasize that these are only *logical* pieces, since all values are still stored in a single array; however, cracking identifies (and incrementally indexes) these pieces and value ranges.

As Figure 3 shows, original cracking reorganizes the array solely based on [low, high], i.e., exactly what the query requested. On the other hand, DDC introduces more knowledge; it first cracks the array on c1, then on c2, and only then on [low, high]. The bound c1 represents the median that cuts the complete array into two pieces with equal number of tuples; likewise, c2 is the median that cuts the left piece into two equal pieces. Thereafter, the newly created piece is found to be small enough; DDC stops searching for medians and *cracks* the piece based on the query's request. For the sake of simplicity, in this example both low and high fall in the same piece and only two iterations are needed to reach a small enough piece size. In general, DDC keeps cutting in half pieces until the minimum allowed size is reached. In addition, the request for [low, high] is evaluated as two requests, one for each bound, as in general each of the two bounds may fall in a different piece.

Figure 4 gives the DDC algorithm. Each query, DDC(C,a,b), attempts to introduce at least two cracks: on $a$ and on $b$ on column $C$. At each iteration, it may introduce (at most $log(N)$) further cracks. Function ddc_crack describes the way DDC cracks for a value $v$. First, it finds the piece that contains the target value $v$ (Lines 4-6). Then, it recursively splits this piece in half while the range of the remaining relevant piece is bigger than CRACK_SIZE (Lines 7-11). Using order statistics, it finds the median $M$ and partitions the array according to $M$ in linear time (Line 9).

For ease of presentation, we avoid the details of the median-finding step in the pseudocode; the general intuition is that we keep reorganizing the piece until we hit the median, i.e., until we create two equal-sized pieces. At first, we simply cut the value *range* in half and try to crack based on the presumed median. Thereafter, we continuously adjust the bounds until we hit the correct median. The median-finding problem is a well-studied problem in computer science, with approaches such as BFPRT [2] providing linear complexity. We use the Introselect algorithm [23], which provides a good worst-case performance by combining quickselect with BFPRT. After the starting piece has been split in half, we choose the half-piece where $v$ falls (Lines 10-11). If that new piece is still large, we keep halving, otherwise we proceed with regular cracking on $v$ and return the final index position of $v$ (Lines 12-13).

In a nutshell, DDC introduces several data-driven cracks until the target piece is small enough. The rationale is that, by halving pieces, we *contain* the cases unfavorable to cracking (i.e., the repeated scans) to small pieces. Thus, the repercussions of such unfavorable cases become negligible. We found that the size of L1 cache as piece size threshold provides the best overall performance.

Still, DDC is also query-driven, as it introduces those cracks only on its path to find the requested values. As seen in Lines 7-11 of Figure 4, it recursively cracks those pieces that contain the requested bound, leaving the rest of the array unoptimized until some other query probes therein. This logic follows the original cracking philosophy, while inseminating it with data-driven elements for the sake of workload robustness. We emphasize that DDC preserves the original cracking interface and column-store requirements; it performs the same task, but adds extra operations therein. As Figure 3 shows, DDC collects all qualifying tuples in a piece of [low, high], as original cracking does.

**Algorithm DDR.** The DDC algorithm introduced several of the core features and philosophy of stochastic cracking, without employing randomness. The type of auxiliary operations employed by DDC are center cracks, always pivoted on a piece's median for optimal partitioning. However, finding these medians is an expensive and data-dependent operation; it burdens individual queries with high and unpredictable costs. As discussed in Section 3, it is critical for cracking, and any adaptive indexing technique, to achieve a low initialization footprint. Queries should not be heavily, if at all, penalized while adapting to the workload. Heavily penalizing a few queries would defeat the purpose of adaptation [10].

Original cracking achieves this goal by performing partitioning and reorganization following only what queries ask for. Still, we have shown that this is not enough when it comes to workload robustness. The DDC algorithm does more than simply following the query's request and thus introduces extra costs. The rest of our algorithms try to strike a good tradeoff between the auxiliary knowledge introduced per query and the overhead we pay for it.

Our first step in this direction is made with the *Data Driven Random* algorithm (DDR), which introduces random elements in its operation. DDR differs from DDC in that it relaxes the requirement that a piece be split exactly in half. Instead, it uses *random* cracks, selecting random pivots until the target value $v$ fits in a piece smaller than the threshold set for the maximum piece size. Thus, DDR can be thought of as a single-branch quicksort. Like quicksort, it splits a piece in two, but, unlike quicksort, it only recurses into one of the two resulting pieces. The choice of that piece is again query-driven, determined by where the requested values fall.

Figure 3 shows an example of how DDR splits an array using initially a random pivot r1, then recursively splits the new left piece on a random pivot r2, and finally cracks based on the requested value range to create piece [low, high]. Admittedly, DDR creates less

**Algorithm** DDC($C, a, b$)
Crack array $C$ on bounds $a, b$.
1. $positionLow$ = ddc_crack($C, a$)
2. $positionHigh$ = ddc_crack($C, b$)
3. $result$ = createView($C, positionLow, positionHigh$)

**function** ddc_crack($C, v$)
4.    Find the piece $Piece$ that contains value $v$
5.    $pLow = Piece$.firstPosition()
6.    $pHgh = Piece$.lastPosition()
7.    **while** ($pHgh - pLow >$ CRACK_SIZE)
8.       $pMiddle = (pLow+pHgh) / 2$;
9.       Introduce crack at $pMiddle$
10.    **if** ($v < C[pMiddle]$) $pHgh = pMiddle$
11.    **else** $pLow = pMiddle$
12.    $position$=crack($C[pLow, pHgh], v$)
13.    $result$=$position$

**Figure 4: The DDC algorithm.**

well-chosen partitions that DDC. Nevertheless, in practice, DDR makes substantially less effort to answer a query, since it does not need to find the correct medians as DDC does, while at the same time it does add auxiliary partitioning information in its randomized way. In a worst-case scenario, DDR may get very unlucky and degenerate to $O(N^2)$ cost; still, it is expected that in practice the randomly chosen pivots will quickly lead to favorable piece sizes.

**Algorithms DD1C and DD1R.** By recursively applying more and more reorganization, both DDC and DDR manage to introduce indexing information that is useful for subsequent queries. Nevertheless, this recursive reorganization may cause the first few queries in a workload to suffer a considerably high overhead in order to perform these auxiliary operations. As we discussed, an adaptive indexing solution should keep the cost of initial queries low [10]. Therefore, we devise two variants of DDC and DDR, which eschew the recursive physical reorganization. These variants perform *at most one* auxiliary physical reorganization. In particular, we devise algorithm DD1C, which works as DDC, with the difference that, after cutting a piece in half, it simply cracks the remaining piece where the requested value is located regardless of its size. Likewise, algorithm DD1R works as DDR, but performs only one random reorganization before it resorts to plain cracking.

DD1C corresponds to the pseudocode description in Figure 4, with the modification that the while statement in Line 7 is replaced by an if statement. Figure 3 shows a high-level example of DD1C and DD1R in action. The figure shows that DD1C cuts only the first piece based on bound c1 and then cracks on [low, high]; likewise, DD1R uses only one random pivot r1. In both cases, the extra steps of their fully recursive siblings are avoided.

**Algorithm MDD1R.** Algorithms DD1C and DD1R try to reduce the initialization overhead of their recursive siblings by performing only one auxiliary reorganization operation, instead of multiple recursive ones. Nevertheless, even this one auxiliary action can be visible in terms of individual query cost, especially for the first query or the first few queries in a workload sequence. That is so because the first query will need to crack the whole column, which for a new workload trend will typically be completely uncracked.

Motivated to further reduce the initialization cost, we devise algorithm MDD1R, where "M" stands for *materialization*. This algorithm works like DD1R, with the difference being that it does *not* perform the final cracking step based on the query bounds. Instead, it materializes the result in a new array.

DD1R and DD1C perform two cracking actions: (1) one for the center or random pivot cracking and (2) one for the query bounds. In contrast, regular cracking performs a single cracking action, only based on the query bounds. Our motivation for MDD1R is to reduce the stochastic cracking costs by eschewing the final cracking operation. Prudently, we do not do away with the random cracking

**Algorithm** MDD1R($C, a, b$)
Crack array $C$ on bounds $a, b$.
1. Find the piece $P1$ that contains value $a$
2. Find the piece $P2$ that contains value $b$
3. **if** ($P1 == P2$)
4.    $result$ = split_and_materialize($P1,a,b$)
5. **else**
6.    $res1$ = split_and_materialize($P1,a,b$)
7.    $res2$ = split_and_materialize($P2,a,b$)
8.    $view$ = createView($C, P1.lastPos+1, P2.firstPos-1$)
9.    $result$ = concat($res1, view, res2$)

**function** split_and_materialize(Piece,a,b)
10. $L$=Piece.firstPosition
11. $R$=Piece.lastPosition
12. $result$=newArray()
13. $X = C[L + rand()\%(R-L+1)]$
14. **while** ($L <= R$)
15.   **while** ($L <= R$ and $C[L] < X$)
16.     **if** ($a <= C[L]$ && $C[L] < b$) $result.Add(C[L])$
17.     $L = L + 1$
18.   **while** ($L <= R$ and $C[R] >= X$)
19.     **if** ($a <= C[R]$ && $C[R] < b$) $result.Add(C[R])$
20.     $R = R - 1$
21.   **if** ($L < R$) **swap**($C[L], C[R]$)
22. Add crack on $X$ at position $L$

**Figure 5: The MDD1R algorithm.**

action, as this is the one that we have introduced aiming to achieve workload robustness. Thus, we drop the cracking action that follows the query bounds. However, we still have to answer the current query (select operator). Therefore, we choose to materialize the result in a new array, just like a plain (non-cracking) select operator does in a column-store. To perform this materialization step efficiently, we integrate it with the random cracking step: we detect and materialize qualifying tuples *while* cracking a data piece based on a random pivot. Otherwise, we would have to do a second scan after the random crack, incurring significant extra cost. Besides, we materialize only when necessary, i.e., we avoid materialization altogether when a query exactly matches a piece, or when qualifying tuples do not exist at the end pieces.

Figure 3 shows high-level view of MDD1R in action. Notably, MDD1R performs the same random crack as DD1R, but does not perform the query-based cracking operation as DD1R does; instead, it just materializes the result tuples. A pseudocode for the MDD1R algorithm is shown in Figure 5.

Figure 6 illustrates a more detailed example on a column that has already been cracked by a number of preceding queries. In general, the two bounds that define a range request in a select operator fall in two different pieces of an already cracked column. MDD1R handles these two pieces independently; it first operates solely on the leftmost piece intersecting with the query range, and then on the rightmost piece, introducing one random crack per piece. In addition, notice that the extra materialization is only *partial*, i.e., the middle qualifying pieces which are not cracked are returned as a view, while only any qualifying tuples from the end pieces need to be materialized. This example also highlights the fact that MDD1R does *not* forgo its query-driven character, even while it eschews query-based cracking per se; it still uses the query bounds to decide *where* to perform its random cracking actions. In other words, the choice of the pivots is random, but the choice of the pieces of the array to be cracked is query-driven.

We do a number of optimizations over the algorithm shown in Figure 5. For example, we reduce the number of comparisons by having specialized versions of the split_and_materialize method. For instance, a request on $[a, b)$ where $a$ and $b$ fall in different pieces, $P1$ and $P2$, will result in two calls, one in $P1$ only, checking for $v > a$, and one on $P2$ only, checking for $v \leq b$.

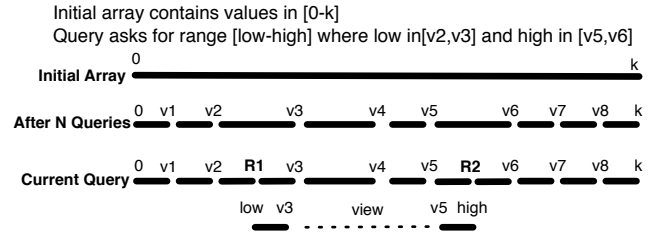

**Figure 6: An example of MDD1R.**

**Progressive Stochastic Cracking.** Our next algorithm, *Progressive* MDD1R (*PMDD1R*) is an even more incremental variant of MDD1R which further reduces the initialization costs. The rationale behind cracking is to build indexes incrementally, as a sequence of several small steps. Each such step is triggered by a single query, and brings about physical reorganization of a column. With PMDD1R we introduce the notion of *progressive* cracking; we take the idea of incremental indexing one step further, and extend it even at the individual cracking steps themselves. PMDD1R completes each cracking operation incrementally, in several partial steps; a physical reorganization action is completed by a sequence of queries, instead of just a single one.

In our design of progressive cracking, we introduce a restriction on the number of physical reorganization actions a single query can perform on a given piece of an array; in particular, we control the number of *swaps* performed to change the position of tuples.

The resulting algorithm is even more lightweight than MDD1R; like MDD1R, it also tries to introduce a single random crack per piece (at most two cracks per query) and materializes part of the result when necessary. The difference of PMDD1R is that it only *gradually* completes the random crack, as more and more queries touch (want to crack) the same piece of the column. For example, say a query $q_1$ needs to crack piece $p_i$. It will then start introducing a random crack on $p_i$, but will only complete part of this operation by allowing $x\%$ swaps to be completed; $q_1$ is fully answered by materializing all qualifying tuples in $p_i$. Then, if a subsequent query $q_2$ needs to crack $p_i$ as well, the random crack initiated by $q_1$, *resumes* while executing $q_2$. Thus, PMDD1R is a generalization of MDD1R; MDD1R is PMDD1R with allowed swaps $x = 100\%$.

We emphasize that the restrictive parameter of the number of swaps allowed per query can be configured as a percentage of the number of tuples in the current piece to be cracked. We will study the effect of this parameter later. In addition, progressive cracking occurs only as long as the targeted data piece is bigger than the L2 cache, otherwise full MDD1R takes over. This provision is necessary in order to avoid slow convergence; we want to use progressive cracking only on large array pieces where the cost of cracking may be significant; otherwise, we prefer to perform cracking as usual so as to reap the benefits of fast convergence.

**Selective Stochastic Cracking.** To further reduce the overhead of stochastic actions, we can *selectively eschew* stochastic cracking for some queries; such queries are answered using original cracking. One approach, which we call *FiftyFifty*, applies stochastic cracking 50% of the time, i.e., only every other query. Still, as we will see, this approach encounters problems due to its deterministic elements, which forsake the robust probabilistic character of stochastic cracking. We propose an enhanced variant, *FlipCoin*, in which the choice of whether to apply stochastic cracking or original cracking for a given query is itself a probabilistic one.

In addition to switching between original and stochastic cracking in a periodic or random manner, we also design a monitoring approach, *ScrackMon*. ScrackMon initiates query processing via original cracking but it also logs all accesses in pieces of a

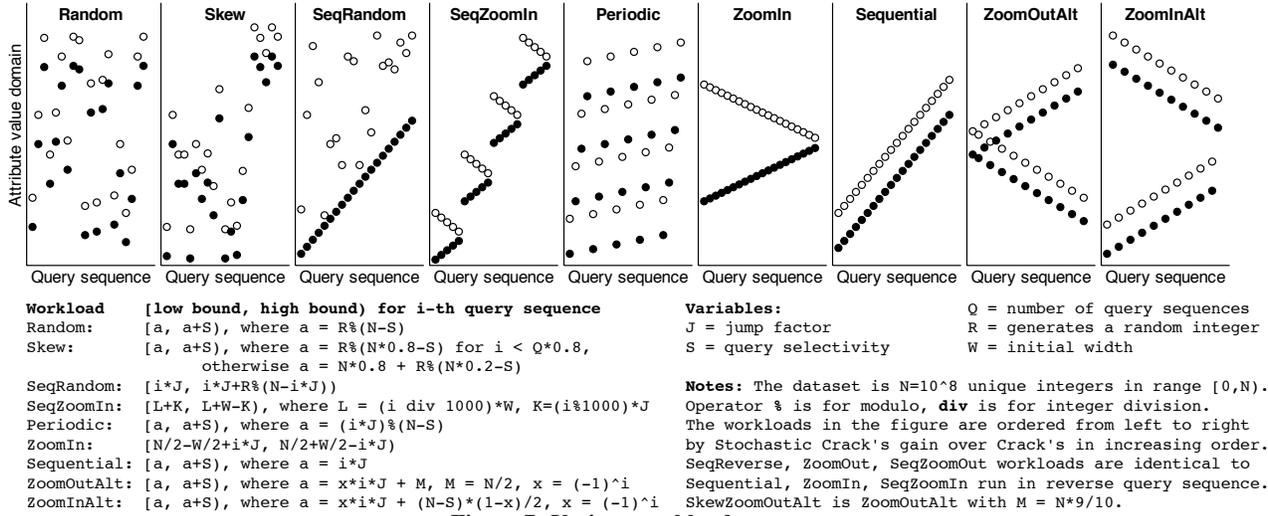

| Workload | [low bound, high bound) for i-th query sequence |
|---|---|
| Random: | [a, a+S), where a = R%(N-S) |
| Skew: | [a, a+S), where a = R%(N*0.8-S) for i < Q*0.8, otherwise a = N*0.8 + R%(N*0.2-S) |
| SeqRandom: | [i*J, i*J+R%(N-i*J)) |
| SeqZoomIn: | [L+K, L+W-K), where L = (i div 1000)*W, K=(i%1000)*J |
| Periodic: | [a, a+S), where a = (i*J)%(N-S) |
| ZoomIn: | [N/2-W/2+i*J, N/2+W/2-i*J) |
| Sequential: | [a, a+S), where a = i*J |
| ZoomOutAlt: | [a, a+S), where a = x*i*J + M, M = N/2, x = (-1)^i |
| ZoomInAlt: | [a, a+S), where a = x*i*J + (N-S)*(1-x)/2, x = (-1)^i |

**Variables:**
J = jump factor
S = query selectivity
Q = number of query sequences
R = generates a random integer
W = initial width

**Notes:** The dataset is N=10^8 unique integers in range [0,N).
Operator % is for modulo, **div** is for integer division.
The workloads in the figure are ordered from left to right by Stochastic Crack's gain over Crack's in increasing order. SeqReverse, ZoomOut, SeqZoomOut workloads are identical to Sequential, ZoomIn, SeqZoomIn run in reverse query sequence. SkewZoomOutAlt is ZoomOutAlt with M = N*9/10.

**Figure 7: Various workloads patterns.**

crack column. Each piece has a crack counter that increases every time this piece is cracked. When a new piece is created it inherits the counter from its parent piece. Once the counter for a piece $p$ reaches a threshold $X$, then the next time ScrackMon uses stochastic cracking to crack $p$, while resetting its counter. This way, ScrackMon monitors all actions on individual pieces and applies stochastic cracking only when necessary and only on problematic data areas with frequent accesses.

Finally, an alternative selective stochastic cracking approach triggers stochastic cracking based on size parameters, i.e., switching from stochastic cracking to original cracking for all pieces in a column which become smaller than L1 cache; within the cache the cracking costs are minimized.

## 5. EXPERIMENTAL ANALYSIS

In this section we demonstrate that Stochastic Cracking solves the workload robustness problem of original cracking.

We implemented all our algorithms in C++, using the C++ Standard Template Library for the cracker indices. All experiments ran on an 8-core hyper-threaded machine (2 Intel E5620 @2.4GHz) with 24GB RAM running CentosOS 5.5 (64-bit). As in past adaptive indexing work, our experiments are all main-memory resident, targeting modern main-memory column-store systems. We use several synthetic workloads as well as a real workload from the scientific domain. The synthetic workloads we use are presented in Figure 7. For each workload, the figure illustrates graphically and mathematically how a sequence of queries touch the attribute value domain of a single column.

**Sequential Workload.** In Section 3, we used the sequential workload as an example of a workload unfavorable for original cracking. We first study the behavior of Stochastic Cracking on the same workload, using exactly the same setup as in Section 3. Figure 9 shows the results. Each graph depicts the cumulative response time, for one or more of the Stochastic Cracking variants, over the query sequence, in logarithmic axes. In addition, each graph shows the plot for original cracking and full indexing (Sort) so as to put the results in perspective. For plain cracking and Sort, the performance is identical to the one seen in Section 3: Sort has a high initial cost and then provides good search performance, while original cracking fails to improve.

**DDC and DDR.** Figure 9(a) depicts the results for DDR and DDC. Our first observation is that both Stochastic Cracking variants manage to avoid the bottleneck that original cracking falls into. They quickly improve their performance and converge to response times similar to those of Sort, producing a quite flat cumulative response time curve. This result demonstrates that, auxiliary reorganization actions can dispel the pathological effect of leaving large portions of the data array completely unindexed.

Comparing DDC and DDR to each other, we observe that DDR carries a significantly smaller footprint regarding its initialization costs, i.e., the cost of the first few queries that carry an adaptation overhead. In the case of DDC, this cost is significantly higher than that of plain cracking (we reiterate that the time axis is logarithmic). This drawback is due to the fact that DDC always tries to find medians and recursively cut pieces into halves. DDR avoids these costs as it uses random pivots instead. Thus, the cost of the first query with DDR is roughly twice faster than that of DDC, and much closer to that of plain cracking.

| | Cumulative time for $10^4$ queries (secs). X=CRACK_AT | | | | |
|---|---|---|---|---|---|
| Workload | X=L1/4 | X=L1/2 | X=L1 | X=L2 | X=3L2 |
| Sequential | 2.2 | 2.2 | 2.2 | 7.8 | 54.7 |

**Figure 8: Varying piece size threshold in DDC.**

In order to demonstrate the effect of the piece size chosen as a threshold for Stochastic Cracking, the table in Figure 8 shows how it affects DDC. L1 provides the best option to avoid cracking actions deemed unnecessary; larger threshold sizes cause performance to degrade due to the increased access costs on larger uncracked pieces. For a threshold even bigger than L2, performance degrades significantly as the access costs are substantial.

**DD1C and DD1R.** Figure 9(b) depicts the behavior of DD1R and DD1C. As with the case of DDR and DDC, DD1R similarly outperforms DD1C by avoiding the costly median search.

Furthermore, by observing Graphs 9(a) and (b), we see that the more lightweight Stochastic Cracking variants (DD1R and DD1C) reduce the initialization overhead compared to their heavier counterparts (DDC and DDR). This is achieved by reducing the number of cracking actions performed with a single query. Naturally, this overhead reduction affects convergence, hence DDR and DDC (Figure 9(a)) converge very quickly to their best-case performance (i.e., their curves flatten) while DD1R and DD1C (Figure 9(b)) require a few more queries to do so (around 10). This extra number of queries depends on the data size; with more data, more queries are needed to index the array sufficiently well.

Notably, DD1R takes slightly longer than DD1C to converge, as it does not always select good cracking pivots, and thus subsequent queries may still have to crack slightly larger pieces than with DD1C. However, the initialization cost of DD1R is about four times less than that of DD1C, whereas the benefits of DD1C are only seen at the point where performance is anyway much better than the worst-case scan-based performance.

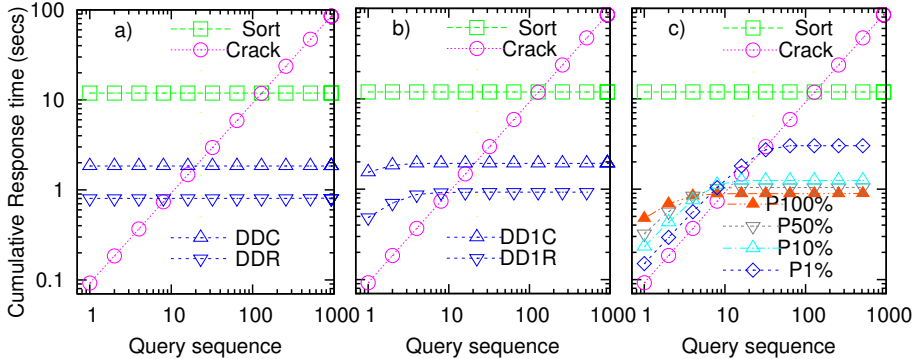
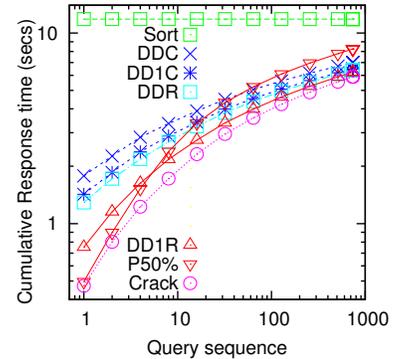

Figure 9: Improving sequential workload via Stochastic Cracking.

Figure 10: Random workload.

**Progressive Stochastic Cracking.** Figure 9(c) depicts the performance of progressive Stochastic Cracking, as a function of the amount of reorganization allowed. For instance, P10% allows for 10% of the tuples to be swapped per query. P100% is the same as MDD1R, imposing no restrictions. The more we constrain the amount of swaps per query in Figure 9(c), the more lightweight the algorithm becomes; thus, P1% achieves a first query performance similar to that of original cracking. Eventually (in this case, after 20 queries), the performance of P1% improves and then quickly converges (i.e., the curve flattens). The other progressive cracking variants obtain faster convergence as they impose fewer restrictions, hence their index reaches a good state much more quickly. Besides, especially in the case of the 10% variant, this relaxation of restrictions does not have a high impact on initialization costs. In effect, by imposing only a minimal initialization overhead, and without a need for workload knowledge or a priori idle time, progressive Stochastic Cracking can tackle this pathological workload.

**Random Workload.** We have now shown that Stochastic Cracking manages to improve over plain cracking with the sequential workload. Still, it remains to be seen whether it maintains the original cracking properties under a random workload as well.

Figure 10 repeats the experiment of Section 3 for the random workload, but adds Stochastic Cracking in the picture (while Scan is omitted for the sake of conciseness). The performance of plain cracking and Sort is as in Section 3; while Sort has a high initialization cost, plain cracking improves in an adaptive way and converges to low response times. Figure 10 shows that *all* our Stochastic Cracking algorithms achieve a performance similar to that of original cracking, maintaining its adaptability and good properties regarding initialization cost and convergence. Moreover, the more lightweight progressive Stochastic Cracking alternative approaches the performance of original cracking quite evenly. Original cracking is marginally faster during the initialization period, i.e., during the first few queries, when the auxiliary actions of Stochastic Cracking operate on larger data pieces, hence are more visible. However, this gain is marginal; with efficient integration of progressive stochastic and query-driven actions, we achieve the same adaptive behavior as with original cracking.

**Varying Selectivity.** The table in Figure 11 shows how Stochastic Cracking maintains its workload robustness with varying selectivity. It shows the cumulative time (seconds) required to run $10^3$ queries. Stochastic cracking maintains its advantage for all selectivities with the sequential workload, while with the random one it adds a bit in terms of cumulative cost over original cracking. However, as shown in Figure 10 (and we will see in Figure 13 for various workloads), this extra cost is amortized across multiple queries and mainly reflects a slightly slower convergence; we argue that this is a rather small price to pay for the overall workload robustness that Stochastic Cracking offers. For example, in Figure 10, DD1R converges to the original cracking performance after 10 queries (in terms of individual response time) while progressive cracking after 20 queries. Going back to the table of Figure 11, we observe that DD1R achieves better cumulative times, while progressive Stochastic Cracking sacrifices a bit more in terms of cumulative costs to allow for a smaller individual query load at the beginning of a workload query sequence (see also Figures 9 and 10). Furthermore, higher selectivity factors cause Scan and progressive cracking to increase their costs, as they have to materialize larger results (whereas the other strategies return non-materialized views as they collect all result tuples in a contiguous area). For progressive cracking, that is only a slight extra cost, as it only has to materialize tuples from the array pieces (at most two) not fully contained within a query's range.

|        | Random Workload selectivity % |           |      |      |      | Sequential Workload selectivity% |           |      |      |      |
|--------|------|------|------|------|------|------|------|------|------|------|
| Algor. | $10^{-7}$ | $10^{-2}$ | 10 | 50 | Rand | $10^{-7}$ | $10^{-2}$ | 10 | 50 | Rand |
| Scan   | 360  | 360  | 500  | 628  | 550  | 125  | 125  | 260  | 550  | 410  |
| Sort   | 11.8 | 11.8 | 11.8 | 11.8 | 11.8 | 11.8 | 11.8 | 11.8 | 11.8 | 11.8 |
| Crack  | 6.1  | 6.0  | 5.7  | 5.9  | 5.9  | 92   | 96   | 108  | 103  | 6    |
| DD1R   | 6.5  | 6.5  | 6.4  | 6.4  | 6.4  | 0.9  | 0.9  | 1.1  | 1.5  | 5.9  |
| P10%   | 8.6  | 8.6  | 10.3 | 10.3 | 10.3 | 1    | 1    | 1.9  | 3.4  | 9.1  |

Figure 11: Varying selectivity.

In the rest of this section, unless otherwise indicated, we use P10% as the default Stochastic Cracking strategy, since we aim at *both* workload robustness *and* low initialization costs.

**Naive Approaches.** A natural question is why we do not simply impose random queries to deal with robustness. The next experiment studies such approaches using the same set-up as before with the sequential workload. In the alternatives shown in Figure 12, R2crack forces 1 random query for every 2 user queries, R4crack forces 1 random query every 4 user queries, and so on. Notably, all these approaches im-prove over original cracking by one order of magnitude in cumulative cost. However, Stochastic Cracking gains another order of magnitude, as it integrates its stochastic cracking actions *within* its query-answering tasks. Furthermore, Stochastic Cracking quickly converges to low response times (its curve becomes flat), while naive approaches do not converge even after $10^3$ queries.

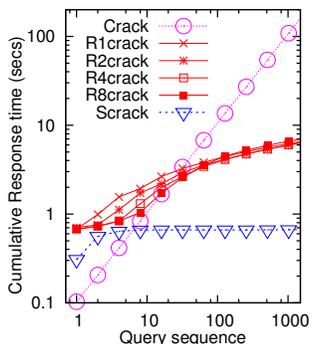

Figure 12: Simple cases.

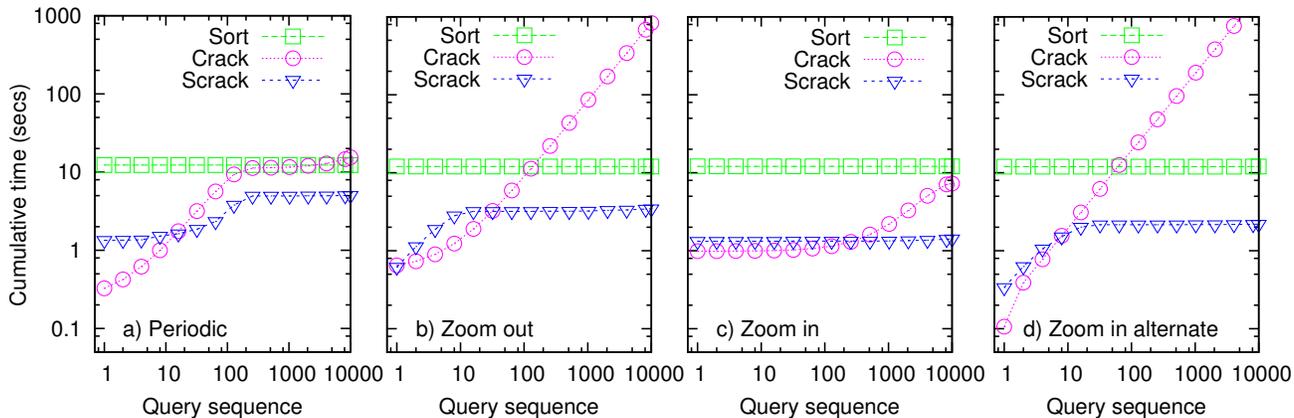
**Figure 13: Various workloads under Stochastic Cracking.**

We conclude that it is preferable to use stochastic cracking algorithms that integrate query-driven with random cracking actions, instead of independently introducing random cracks. This rationale is the same as that in original cracking: physical refinement is not an "afterthought", an action merely triggered by a query; it is *integrated* in the query processing operators and occurs on the fly.

**Adaptive Indexing Hybrids.** In recent work, cracking was extended with a partition/merge logic [19]. Therewith, a column is split into multiple pieces and each piece is cracked independently. Then, the relevant data for a query is merged out of all pieces. These partition/merge-like algorithms improve over original cracking by allowing for better access patterns. However, as they are still based on what we call the blinkered query-driven philosophy of original cracking, they are also expected to suffer from the kind of workload robustness problems that we have observed. Figure 14 demonstrates our claim, using the sequential workload. We use the Crack-Crack (AICC) and Crack-Sort (AICS) methods from [19]. They both fail to improve on their performance, as they blindly follow the workload. Besides, due to the extra merging overhead imposed by the sequential workload, AICC and AICS are both slightly slower than original cracking. In order to see the effect and application of Stochastic Cracking in this case as well, we implemented the basic stochastic cracking logic inside AICS and AICC, in the same way we did for DD1R. The same figure, above, shows the performance of AICS1R and AICC1R, namely our algorithms, which, in addition to the cracking and partition/merge logic, also incorporate DD1R-like stochastic cracking in one go during query processing. Both our stochastic cracking variants gracefully adapt to the Sequential Workload, quickly converging to low response times. Thereby, we demonstrate that the concept of stochastic cracking is directly applicable and useful to the core cracking routines, wherever these may be used.

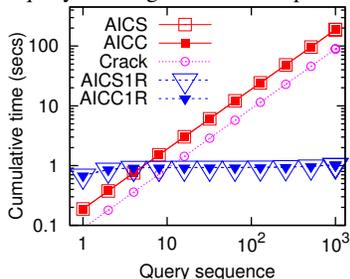
**Figure 14: Stochastic hybrids.**

**Various Workloads.** Next, we compare Stochastic Cracking against original cracking and full indexing (Sort) on a variety of workloads. Figure 13 shows the results with 4 of the workloads from Figure 7. Stochastic cracking performs robustly across the whole spectrum of workloads. On the other hand, original cracking fails in many cases; in half of the workloads, it loses the low initialization advantage over full indexing, and performs significantly worse than both Stochastic Cracking and full indexing over the complete workload. In all these workloads, the order in which queries are posed forces cracking on subsequent queries to deal with a large data area all over again. At the same time, for those workloads where original cracking does not fail, Stochastic Cracking follows a similar behavior and performance. Only for the random workload is original cracking marginally faster over Stochastic Cracking, gaining 1.4 seconds over the course of $10^4$ queries.

**Updates.** Figure 15 on the left shows the Stochastic Cracking performance under updates. Given that stochastic cracking maintains the core cracking architecture, the update techniques proposed in [17] apply here as well. Updates are marked and collected as *pending* updates upon arrival. When a query $Q$ requests values in a range where at least one pending update falls, then the qualifying updates for the given query are merged during cracking for $Q$. We use the Ripple algorithm [17] to minimize the cost of merging, i.e., reorganizing dense arrays in a column-store. The figure presents the performance with the Sequential workload when updates interleave with queries. We test a high-frequency update scenario where 10 random updates arrive every with 10 queries. Notably, Stochastic Cracking maintains its advantages and robust behavior, not being affected by updates. We obtained the same behavior with varying update frequency (as in [17]).

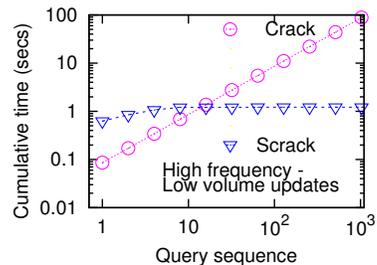
**Figure 15: Adaptive updates.**

**Stochastic Cracking on Real Workloads.** In our next experiment, we test Stochastic Cracking on a SkyServer workload [25]. The SkyServer contains data from the astronomy domain and provides public database access to individual users and institutions. We used a 4 Terabyte SkyServer data set. To focus on the effect of the select operator, which matters for Stochastic Cracking, we filtered the selection predicates from queries and applied them in exactly the same chronological order in which they were posed in the system. Figure 16(b) depicts the exact workload pattern logged in the SkyServer for queries using the "right ascension" attribute of the "Photoobjall" table. The Photoobjall table contains 500 million tuples, and is one of the most commonly used ones. Overall, we observe that all users/institutions pose queries following non-random patterns. The queries focus in a specific area of the sky before moving on to a different area; the pattern combines features of the synthetic workloads we have studied. As with those workloads, here too, the fact that queries focus on one area at a time creates large unindexed areas. Figure 16(a) shows that plain crack-ing fails to provide robustness in this case as well, while Stochastic Cracking maintains robust performance throughout the query sequence; it answers all 160 thousand queries in only 25 seconds, while original cracking needs more than 2000 seconds (a full indexing approach needs 70 seconds, while a plain scan more than 8000 seconds). These results establish the advantage of Stochastic Cracking with a real workload.

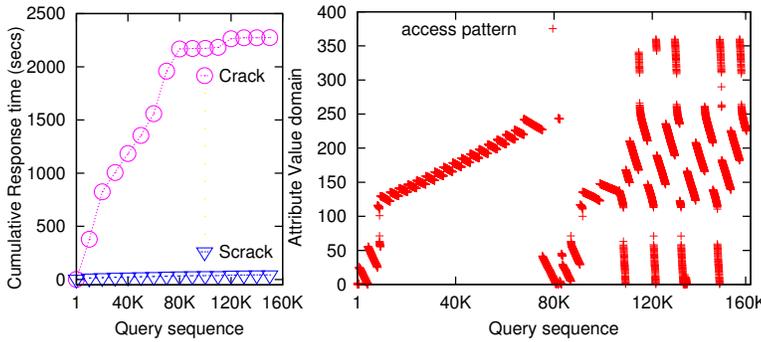

**Figure 16: Cracking on the SkyServer workload.**

| Workload | Cracking strategy (secs) | | | |
|---|---|---|---|---|
| | Crack | Scrack | FiftyFifty | FlipCoin |
| Periodic | 15.4 | 5 | 8.4 | 6.9 |
| ZoomOut | 1019 | 1.6 | 2 | 2 |
| ZoomIn | 7.2 | 1.4 | 1.3 | 2 |
| ZoomInAlt | 1822 | 1.8 | 916 | 1.2 |
| Random | 8.6 | 10 | 9.5 | 9.4 |
| Skew | 7.6 | 7.1 | 8.8 | 8.7 |
| SeqReverse | 2791 | 1 | 1.8 | 1.6 |
| SeqZoomIn | 2.3 | 1.2 | 1.9 | 1.2 |
| SeqRandom | 8.6 | 9.6 | 7.8 | 9.2 |
| Sequential | 861 | 0.4 | 1.6 | 2.4 |
| SeqZoomOut | 1215 | 1.3 | 2 | 1.5 |
| ZoomOutAlt | 920 | 1.2 | 224 | 1.2 |
| SkewZoomOutAlt | 1382 | 1.1 | 1381 | 2.2 |
| Mixed | 331 | 3.2 | 30.5 | 4.5 |
| SkyServer | 2274 | 25 | 62 | 35 |

**Figure 17: Various workloads.**

| Workload | Strategy, Cumulative time for $1.6 * 10^5$ queries (secs) Stochastic crack every $X$ queries, original crack otherwise | | | | | |
|---|---|---|---|---|---|---|
| | X=1, (Scrack) | X=2, (FiftyFifty) | X=4 | X=8 | X=16 | X=32 |
| SkyServer | 25 | 62 | 65 | 97 | 153 | 239 |

**Figure 18: Selective Stochastic Cracking with a varying period.**

**Selective Stochastic Cracking.** In our last experiment, we evaluate the potential of Selective Stochastic Cracking to further optimize Stochastic Cracking. Figure 17 presents the cumulative time to run $10^4$ queries under various workloads. It first shows results for the 4 workloads shown in Figure 13, and then for an additional set of extra workloads to cover all workloads defined in Figure 7. In addition to individual workload patterns, Figure 17 also depicts results for a Mixed workload representing a mixture of all workloads studied so far; it randomly switches between each workload in every 1000 queries. Furthermore, Figure 17 shows the performance on the SkyServer workload ($1.6 * 10^5$ queries). All Stochastic Cracking variants use MDD1R, which provides a good balance of initialization costs vs. cumulative run time performance.

First, we observe that Stochastic Cracking maintains its robust behavior across all new workloads. On the other hand, original cracking fails significantly with most of them, being two or more orders of magnitude slower than Stochastic Cracking. Original cracking behaves well only for the workloads that contain enough random elements by themselves; even then, its benefit over Stochastic Cracking is only 1 second over the course of $10^4$ queries.

Comparing Stochastic Cracking with its Selective variant, we observe that FiftyFifty behaves rather well in many scenarios, but still fails in some of them, i.e., it is not robust. This is due to the fact that it follows a query-driven logic with every second query; thus, it is vulnerable to patterns that happen to create big column pieces during (some of the) odd queries.

If we decrease the frequency with which we apply Stochastic Cracking, then the performance degrades even further. The table in Figure 18 demonstrates this effect on the SkyServer workload ($1.6 * 10^5$ queries). As we apply less Stochastic Cracking, Selective Stochastic Cracking becomes increasingly more likely to take bad decisions when applying original cracking. Thus, Continuous Stochastic Cracking ($X$=1) outperforms all its Selective variants, as consistently uses stochastic elements.

Coming back to Figure 17, we see that the FlipCoin strategy provides an overall robust solution, i.e., it does not fail in any of the workloads. By randomizing the decision on whether to apply Stochastic Cracking or not for every query, it avoids the deterministic bad access patterns that may appear with each workload. In the SkyServer workload, FlipCoin needed 35 seconds as opposed to 62 seconds for FiftyFifty. However, its performance remains at a disadvantage when compared to pure Stochastic Cracking (25 seconds). Again, this is due to the fact that FlipCoin may fall into bad access patterns (even if only a few), as it eschews stochastic operations where it should not. Thus, out of these two Selective Stochastic Cracking variants, neither the deterministic FiftyFifty, nor the probabilistic FlipCoin, manages to present an overall better performance than pure Stochastic Cracking.

| Workload | Strategy, Cumulative time for $1.6 * 10^5$ queries (secs) Continuous monitoring: use stochastic crack in a piece $P$ when $P.CrackCounter = X$, otherwise original crack | | | | | |
|---|---|---|---|---|---|---|
| | X=1, (Scrack) | X=5 | X=10 | X=50 | X=100 | X=500 |
| SkyServer | 25 | 83 | 127 | 366 | 585 | 1316 |

**Figure 19: Selective Stochastic Cracking via monitoring.**

Figure 19 shows how Selective Stochastic Cracking via monitoring behaves on the SkyServer workload. This approach treats each cracking piece *independently* and applies stochastic actions directly on problematic pieces; it can detect bad access patterns even if they occur on isolated pieces as opposed to the whole column. Similarly to what we observed for other Selective Cracking approaches, as we increase the monitoring threshold, i.e. the number of queries we allow to touch a piece until we trigger Stochastic Cracking therefor, the more performance degrades. Again, Continuous Stochastic Cracking, i.e., Stochastic Cracking applied with every access on a column piece, provides the best performance.

We also experimented with approaches that stop applying stochastic cracking when a piece becomes smaller than the L1 cache; it then resorts to original cracking. Such approaches turned out to create a significant overhead, i.e., performance 2-3 times slower than pure Stochastic Cracking across all workloads, except for the Random one where they improve by 10-20%. If we increase the piece size threshold, performance degrades further (cf. our earlier observations when varying the DDC piece threshold).

**Summary.** We have shown that original cracking relies on the randomness of the workloads to converge well. However, where the workload is non-random, cracking needs to introduce randomness on its own. Stochastic Cracking clearly improves over original cracking by being robust in workload changes while maintaining all original cracking features when it comes to adaptation. Furthermore, we have established that, given the unpredictability of dynamic workloads, there is no "royal road" to workload robustness, i.e., no easy way out of the necessity to apply stochastic cracking operations with *every single query*. This consistent strategy provides an overall robust behavior with non-random workloads, while raising only a minimal overhead with random ones.

Our final graph in Figure 20 summarizes the performance of the Stochastic Cracking variants using the Sequential workload. The $x$-axis shows the total cumulative time to run the whole query sequence. The $y$-axis represents the cumulative cost to run the first few queries, i.e., the first one, two, four, etc. In other words, the more lightweight algorithms in terms of total cost appear in the leftmost part of the $x$-axis, while the more lightweight algorithms in terms of initialization cost (first few queries) are the ones whose line appears lower along the $y$-axis. For the sake of readability, we present only three representative variants. DD1R proves to be the best choice regarding total costs, while progressive cracking can be tuned depending on the desired behavior; best initialization costs versus overall cumulative costs.

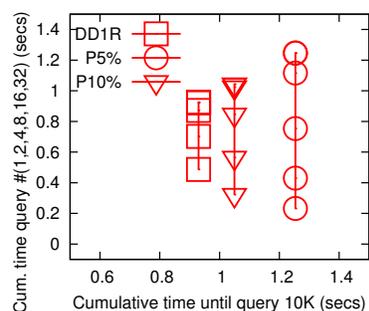

Figure 20: Summary.

## 6. FUTURE WORK

For future work, we can distinguish between (a) open topics for cracking in general and (b) optimizations for stochastic cracking.

**Database Cracking.** Open topics for cracking include concurrency control and disk-based processing. The first steps towards this direction have been done in [19] and [12]. The challenge with concurrent queries is that the physical reorganizations they incur have to be synchronized, possibly with proper fine grained locking. Disk-based processing poses a challenge because the continuous reorganization may cause continuous writes to disk; we need to examine how much reorganization we can afford per query without increasing I/O costs prohibitively. Other open topics include distribution, compression, optimization, maintenance, as well as performance studies across different storage mediums such as flash disks, memory, caches, and hard disks.

**Stochastic Cracking.** Crucially, stochastic cracking does not violate the design principles and interfaces of original cracking. The algorithmic changes are local within the basic physical reorganization algorithms and thus do not cause any undesired side-effects to the design of a cracking kernel at large. For future work, we identify the further reduction of the initialization cost to make stochastic cracking even more transparent to the user, especially for queries that initiate a workload change and hence incur a higher cost. Another line of improvement lies in combining the strengths of the various stochastic cracking algorithms via a dynamic component that decides which algorithm to choose for a query on the fly.

## 7. CONCLUSIONS

This paper introduced Stochastic Database Cracking, a proposal that solves the workload robustness deficiency inherent in Database Cracking as originally proposed. Like original cracking, Stochastic Cracking works adaptively and incrementally, with minimal impact on query processing. At the same time, it solves a major open problem, namely the sensitivity of the cracking process to the kind of queries posed. We have shown that the effectiveness of original cracking deteriorates under certain workload patterns due to its strict reliance on the workload for extracting indexing hints. Stochastic Cracking alleviates this problem by introducing random physical reorganization steps for efficient incremental index-building, while also taking the actual queries into account. We proposed several Stochastic Cracking variants; with thorough experimentation, we have demonstrated that Stochastic Cracking expands the benefits of original cracking to a much wider variety of workloads. Our Progressive Stochastic Cracking method carries the cracking methodology to a new formulation, as it shares even a single index-refinement operation among several queries in order to amortize its cost. Moreover, we have shown that a lighter, partial, or occasional application of stochastic operations does not bring forth equally good results. Overall, our work renders database cracking effective in a much wider variety of scenarios.

## Acknowledgements

We are grateful to Lefteris Sidirourgos for insights regarding the SkyServer experiments. We also thank the anonymous reviewers for their constructive feedback.